\documentclass[aps,preprint,epsfig,rotate]{revtex4}
\begin{document}
\title{On the isotopic shifts in the light two-electron atoms and ions.}

 \author{Alexei M. Frolov}
 \email[E--mail address: ]{afrolov@uwo.ca}

\affiliation{Department of Chemistry\\
 University of Western Ontario, London, Ontario N6H 5B7, Canada}

\date{\today}

\begin{abstract}

The isotopic shifts are determined to high accuracy for a number of light 
two-electron Li$^{+}$, Be$^{2+}$, B$^{3+}$ and C$^{4+}$ ions. The field 
components of these isotopic shifts have been determined with the use of 
the exact Racah-Rosental-Breit formula. 

\end{abstract}

\maketitle
\newpage

The dependence of the total energies of bound states of different isotopes of the 
same chemical element upon the `isotope numbers' is called the isotopic shift. By the
`isotope numbers' we mean the mass of the nucleus and its electric charge density 
distribution (or proton density distribution). In general, the isotopic shift can be 
observed in arbitrary, in principle, atomic system which has a central, heavy nucleus 
with the finite mass, positive electric charge and finite radius. In general, the 
isotopic shift $\Delta E$ of the bound state level with the total energy $E$ has the 
two following components: the first component which depends upon the mass of the 
nucleus and the second component which mainly depends upon the electric charge 
density distribution in the atomic nucleus. In turn, the first component is 
represented as the sum of the normal and specific components. Each of these two 
components is proportional to the factor $\frac{m_e}{M}$, where $m_e$ is the mass of 
the electron, while $M$ is the nuclear mass. For few-electron ($N-$electron) atoms 
and ions the exact formula for the isotopic shift $\Delta E$ takes the form
\begin{eqnarray}
 \Delta E = \Delta E_{n} + \Delta E_{s} = \frac{m_e}{M} \langle \sum^{N}_{i=1} 
 \frac{{\bf p}^2_i}{2 m_e} \rangle + \frac{m_e}{M} \langle \sum^{N}_{i (i \ne k) = 1} 
 \frac{{\bf p}_i \cdot {\bf p}_k}{2 m_e} \rangle  \label{eq1}
\end{eqnarray}
where the notation $\langle \hat{X} \rangle$ designates the expectation value of the
operator $\hat{X}$. For the two-electron (or helium-like) atoms and ions in atomic 
units $\hbar = 1, m_e = 1, e = 1$ the last formula is reduced to the form
\begin{eqnarray}
 \Delta E = \Delta E_{n} + \Delta E_{s} = \frac{1}{M} \langle {\bf p}^2_1 \rangle + 
 \frac{1}{2 M} \langle {\bf p}_1 \cdot {\bf p}_2 \rangle  \label{eq2}
\end{eqnarray}
As follows from this formula the expression for the normal and specific components of 
the isotopic shift are 
\begin{eqnarray}
 \Delta E_{n} = \frac{1}{M} \langle {\bf p}^2_1 \rangle \; \; \; and \; \; \; 
 \Delta E_{s} = \frac{1}{2 M} \langle {\bf p}_1 \cdot {\bf p}_2 \rangle \; \; \; , 
 \label{eq3}
\end{eqnarray}
respectively. As follows from Eq.(\ref{eq3}) to determine the normal and specific 
components of the isotopic shift in two-electron atom/ion one needs to obtain the 
expectation values of the ${\bf p}^2_1$ and ${\bf p}_1 \cdot {\bf p}_2$ operators. 
Everywhere in this study we assume that the wave functions of the two-electron 
atom/ion are properly symmetrized upon spin-spatial permutations of the two electrons 
and, therefore, the corresponding single-electron expectation values are always equal 
to each other, e.g., $\langle {\bf p}^2_1 \rangle = \langle {\bf p}^2_2 \rangle$.

In actual two-electron atomic systems, i.e. in the systems with the finite nuclear 
mass $M$, one can use the condition which follows from the conservation of the total 
momentum ${\bf P}_N = {\bf p}_1 + {\bf p}_2$, where ${\bf P}_N$ is the momentum of 
the nucleus, while ${\bf p}_1$ and ${\bf p}_2$ are the electron momenta. From here
one finds:   
\begin{eqnarray}
 \frac12 \langle {\bf P}^2_N \rangle = \langle {\bf p}^2_1 \rangle + 
 \langle {\bf p}_1 \cdot {\bf p}_2 \rangle  \label{eq4}
\end{eqnarray}
and, therefore, from Eqs.(\ref{eq2}) and (\ref{eq4}) we have
\begin{eqnarray}
 \Delta E = \frac{1}{2 M} \langle {\bf P}^2_N \rangle \label{eq5}
\end{eqnarray}
i.e. the mass dependent component of the isotopic shift is the expectation value of 
the kinetic energy of the atomic nucleus with the large (but finite!) mass. In many 
books and textbooks the formula, Eq.(\ref{eq5}), is considered as the original (or 
fundamental) expression, while Eq.(\ref{eq1}) is derived from this formula.  

As mentioned above the field component of the isotopic shift explicitly depends upon 
the nuclear size (radius) $R$ and proton distribution density in the nucleus. It is 
clear that this component also depends upon the mass of atomic nucleus $M$, since the 
nuclear matter is a saturated matter (in contrast with the Coulomb matter). Therefore, 
the nuclear radius $R$ is uniformly related to the number of nucleons $A$ in the 
nucleus: $R = r_0 \cdot A^{\frac13}$, where the `constant' radius $r_0 \approx 1.17 - 
1.25 \cdot 10^{-13}$ $cm$ = 1.17 - 1.25 $fm$ (fermi), where 1 $fm$ = $1 \cdot 10^{-13}$ 
$cm$. Briefly, this means that the field component of the total isotopic shift is also 
a function of the nuclear mass $M$, since $A \approx \frac{M}{m_p}$, where $m_p$ is the 
proton mass. The more accurate formula for the nuclear mass as the function of $A, Z (= 
N_p)$, where $Z$ is the electric charge of the nucleus = number of protons $N_p$) and 
$N_n$ (number of neutrons) is given by the Weiz\"{a}cker formula. This formula is 
discussed in the Appendix. 

Let us present here the well known formula obtained by Racah, Rosental and Breit for 
the field shift (see, e.g., \cite{Sob} and references therein) 
\begin{eqnarray}
 E^{fs}_M = \frac{4 \pi a^2_0}{Q} \cdot \frac{b + 1}{[\Gamma(2 b + 1)]^2} \cdot B(b) \cdot
 \Bigl( \frac{2 Q R}{a_0} \Bigr)^{2 b} \cdot \frac{\delta R}{R} \cdot 
 \langle \delta({\bf r}_{eN}) \rangle \label{eqf3}
\end{eqnarray}
where $Q$ is the nuclear charge, $R$ is the nuclear radius and $b = \sqrt{1 - \alpha^2 
Q^2}$, where $\alpha = \frac{e^2}{\hbar c} \approx \frac{1}{137}$ is the dimensionless
constant which is the small parameter in QED. In Eq.(\ref{eq3}) the notation $\Gamma(x)$ 
stands for the Euler's gamma-function, while the factor $B(b)$ is directly related to the 
proton density distribution in the atomic nucleus. By assuming the uniform distribution 
of the proton density over the volume of the nucleus one finds the following expression 
for the factor $B(b)$
\begin{eqnarray}
 B(b) = \frac{3}{(2 b + 1) (2 b + 3)} \label{eqf4} 
\end{eqnarray}
For light nuclei with $Q \le 6$ we have $b \approx 1$ and $B \approx \frac15$. The 
formula, Eq.(\ref{eqf3}), was used in many theoretical papers for numerical evaluations 
of the filed component of the isotopic shift, or field shift for short. In some works, 
however, this formula was written with a number of `obvious simplifications'. Many 
of such `simplifications' are based on the fact that for light nuclei the numerical 
value of the factor $b$ is close to unity. Furthermore, in some papers the factor $b$ 
was mistakenly called and considered as the Lorentz factor, while the actual Lorentz 
factor $\gamma$ is its inverse value $\gamma = \frac{1}{b} = \frac{1}{\sqrt{1 - 
\alpha^2 Q^2}}$. Such a factor $\gamma$ always exceeds unity.

In this study we evaluate the isotopic shift for a number of the ground $1^1S(L = 
0)-$states in the light two-electron ions by using the exact (not approximate!) formula, 
Eq.(\ref{eqf3}). Such evaluations allow one to evaluate the numerical errors which arise 
from the use of approximate expressions. Our main interest in this study is related with 
the field component of the isotopic shift. As follows from Eq.(\ref{eqf3}) to evaluate 
the field component of the field shift one needs to determine to very high accuracy the 
expectation value of the electron-nuclear delta-function (or the $\langle \delta({\bf 
r}_{eN}) \rangle$ value). In this study for each light element/atom all isotopic shifts 
are determined in respect to the model (or idealized) isotope which has the infinitely 
heavy nucleus with zero spatial radius (zero charged radius). In this case the ratio 
$\frac{\delta R}{R}$ in  Eq.(\ref{eqf3}) equals unity, and this equation can be 
re-written to the form (in atomic units)
\begin{eqnarray}
 E^{fs}_M = 4^{b+1} \pi Q^{2b-1} \cdot \alpha^{4 b} \cdot \frac{3 (b + 1)}{[\Gamma(2 
 b + 1)]^2 (2 b + 1) (2 b + 3)} \cdot \Bigl( \frac{R}{r_e} \Bigr)^{2 b} \cdot 
 \langle \delta({\bf r}_{eN}) \rangle \label{eqf5}
\end{eqnarray}
where $r_e = \alpha^2 a_0 \approx 2.817940$ $fm$ (1 $fm$ = $1 c\cdot 10^{-13}$ $cm$)) is 
the classical radius of the electron. For light atomic nuclei the dimensionless factor 
$\frac{R}{r_e}$ in the last formula is close to unity. Also, in our calculations we have
used the following numerical values for the physical constants: $\alpha = 7.297352569 
\cdot 10^{-3}$ and $a_0 = 5.291772109 \cdot 10^{-9}$ $cm$. The formula, Eq.(\ref{eqf5}), 
has been used in all calculations performed for this study. As follows from Eq.(\ref{eqf5}) 
to determine the field component of the isotopic shift one needs to know the expectation 
value of the electron-nuclear delta-function $\delta({\bf r}_{eN})$ and numerical value of 
the nuclear radius $R$. The expectation value of $\delta({\bf r}_{eN})$ can be found from 
the results of highly accurate atomic computations, while the nuclear radius must be 
taken from nuclear experiments (see, e.g., \cite{Angel}).

In this study we consider a few light two-electron (or He-like) ions: Li$^{+}$, 
Be$^{2+}$, B$^{3+}$ and C$^{4+}$. The total energies and some other properties (or 
expectation values) of the corresponding model ions (with the infinite nuclear mass 
and zero spatial radii) can be found in Tables I - IV. The convergence of the 
results upon the total number of basis function $N$ used in calculations is also shown 
in Tables I - IV. For these Tables we have computed the expectation values of the
following operators: $\delta({\bf r}_{eN}), \nu_{eN}, {\bf p}^2_1, {\bf p}_1 \cdot 
{\bf p}_2$ and ${\bf p}^2_N$. The notation $\nu_{eN}$ designates the electron-nuclear 
cusp value which is defined by the following equation
\begin{eqnarray}
 \nu_{eN} = \frac{\langle \delta({\bf r}_{eN}) \frac{\partial}{\partial r_{eN}} 
 \rangle}{\langle \delta({\bf r}_{eN}) \rangle} \label{eqf6}
\end{eqnarray}
Formally, it is an avaraged velocity of the electron at the nucleus. In the general 
case, this expectation value gives the relative velocity of the two particles ($i$ and 
$j$) at the $(ij)-$collision point. For pure Coulomb systems, e.g., for the atom with 
the nuclear charge $Q$, such a velocity is known from the corresponding classical 
problem. In particular, for the atom with the nuclear charge $Q$, the electron-nuclear 
cusp $\nu_{eN}$ must be equal (in atomic units)
\begin{eqnarray}
 \nu_{eN} = - Q e^2 \frac{m_e M_N}{m_e + M_N} = - Q \label{eqf7}
\end{eqnarray}
since $M_N = \infty$.

The coincidence of the computed $\nu_{eN}$ value with the nuclear charge $Q$ (or -$Q$) 
indicates the quality of the variational (atomic) wave functions around the nucleus and 
overall accuracy of the computed expectation value of the electron-nuclear delta-function. 
In all calculations performed for this study we have used our exponential variational 
expansion in relative coordinates described in detail in our earlier papers, see, e.g., 
\cite{Fro01}, \cite{Fro06}. Here we do not want to repeat these descriptions of our 
variational expansion. Note also that in our calculations we have determined many dozens 
of different expectation values, including singular expectation values. However, in 
Tables I - IV only to a few expectation values are presented. All these values are needed 
for numerical evaluation of the isotopic shifts. 

Table V contains the numerical values of the field components of isotopic shifts (in 
$a.u.$) derermined with the use of the formula Eq.(\ref{eqf5}). In this Table we also 
present the numerical values of the following factors from that formula: $R$ (the actual 
nuclear radius), $b, X = 4^{b+1} \pi Q^{2b-1} \cdot \alpha^{4 b} \cdot \frac{3 (b + 
1)}{[\Gamma(2 b + 1)]^2 (2 b + 1) (2 b + 3)}$ and $Y = \Bigl( \frac{R}{r_e} \Bigr)^{2 
b}$. The expectation values of the electron-nuclear delta-functions were taken from 
Tables I - IV. To evaluate the Euler's gamma-function $\Gamma(x)$ we have used the 
approximate 7-term formula derived by Lanczos \cite{Lanc}. The overall accuracy of this
formula is $\approx 1 \cdot 10^{-10} - 2 \cdot 10^{-10}$.    

\begin{center}
 {\bf Appendix I.}
\end{center}

The formula which provides the uniform relation between the nuclear mass $M$ and total 
number of nucleons $A$, nuclear charge $Z$ (= number of protons $N_p$) and number of 
neutrons $N_n$ in the nucleus was derived in 1937 by Bethe, Weiz\"{a}cker and others. 
Now, this formula is know as the Weiz\"{a}cker formula \cite{Weiz}, or Bethe-Weiz\"{a}cker 
formula. This five-term formula for the nuclear binding energy $E_b$ was produced 75 
years ago and since then its general structure has never been changed. First, note that
the mass formula for an arbitrary nucleus is written in the form 
\begin{eqnarray}
 M = m_p \Bigl[ Z + N \Bigl( \frac{m_n}{m_p} \Bigr) - \frac{E_b}{m_p c^2} \Bigr] 
 \label{eap1}
\end{eqnarray}
where $M$ is the nuclear mass of the nucleus with $A$ nucleons, $Z$ protons and $N$
neutrons $A = Z + N$). Also in this formula $E_b$ is the binding energy of the nucleus,
$c$ is the speed of light in vacuum, while $m_p$ and $m_n$ are the masses of the proton 
and neutron, respectively. The factors $m_p c^2$ = 938.272910 $MeV$ and $m_n c^2$ = 
939.565378 $MeV$. The advantage of the formula, Eq.(\ref{eap1}), is obvious, since it 
contains only dimensionless ratios and two integer numbers ($Z$ and $N$). For instance, 
if we chose in Eq.(\ref{eap1}) $m_p = 1836.152701 m_e$, then $M$ will be given in $m_e$ 
(or in atomic units if $m_e = 1$). This is very convenient for highly accurate 
computations of different few-electron ions. 

The parameter $E_b$ in Eq.(\ref{eap1}) is called the binding energy of the nucleus. The 
explicit expression for the nuclear binding energy $E_b$ is written as the following sum 
(the Weiz\"{a}cker formula):
\begin{eqnarray}
 E_b = a_V A - a_S A^{\frac23} - a_C \frac{Z^2}{A^{\frac13}} - a_A \frac{(N - Z)^2}{A}
 + \delta(A,Z) \label{eap2}
\end{eqnarray}
where the five terms in the right-hand side of this equation are called the volume term,
surface term, Coulomb term, assymetry term and pairing term, respectively. The pairing
term $\delta(A,Z)$ equals zero, if $A$ is odd. If $A$ is even and both $Z$ and $N$ are 
even, then $\delta(A,Z) = \frac{a_p}{\sqrt{A}}$. The Weiz\"{a}cker formula is relatively
accurate for regular nuclei (i.e. for nuclei which are not far from the stability 
region). In reality, such an accuracy directly depends upon the numerical values of 
parameters $a_V, a_S, a_C, a_A$ and $a_p$ in Eq.(\ref{eap2}). In our calculations we have
used the following values of these parameters: $a_V$ = 15.8 $MeV$, $a_S$ = 18.3 $MeV$, 
$a_C$ = 0.714 $MeV$, $a_A$ = 23.2 $MeV$ and $a_p$ = 12.0 $MeV$. The Weiz\"{a}cker formula
with these coefficients is sufficiently accurate for all light nuclei which are located in 
the stability region. 

\begin{center}
 {\bf Appendix II.}
\end{center}

In general, the bound state in any few-electron atom arises in the result of proper 
balance between kinetic energy $T_e$ of atomic electrons, electron-nuclear (Coulomb) 
attraction $U_{en}$ and electron-electron (Coulomb) repulsion $U_{ee}$. The total 
energy $E$ of this bound state is written as the following sum:
\begin{eqnarray}
   E = T_e + U_{ne} + U_{ee} =  \frac12 (U_{ne} + U_{ee}) \label{eq1x}
\end{eqnarray}
where in the right-hand side we have used the virial theorem for the Coulomb potential
(see, e..g, \cite{Fock}) $T_e = -\frac12 (U_{ne} + U_{ee})$. To solve the problem we need
the explicit expression for the correlation energy $U_{ee}$ in terms of the electron-nuclear
attraction energy $U_{ne}$. Such an analytical relation can be found in a few restricted 
cases, e.g., in the Thomas-Fermi method one finds: $U_{ee} = - \frac17 U_{ne}$ and $E = 
\frac37 U_{ne}$. For highly accurate method the situation is significantly more complicated,
but qualitatively the actual relations between $T_e, U_{ne}$ and $U_{ee}$ are similar to the 
relations obeyed in the Thomas-Fermi model.
 
First, note that the Schr\"{o}dinger equation for an arbitrary non-relativistic $N-$electron 
atom/ion takes the form 
\begin{eqnarray}
 \Bigl[\frac{\hbar^2}{2 m_e} \sum^{N}_{i=1} {\bf p}^2_i - Q e^2 \sum^{N}_{i=1} \frac{1}{r_i} 
 + e^2 \sum^{N}_{i>j=1} \frac{1}{r_{ij}} \Bigr] \Psi = E \Psi \label{eq2x}
\end{eqnarray}
where $Q$ is the electric charge of the nucleus and the mass of the central nucleus is 
assumed to be finite. Here we shall not assume that $N = Q$. In Eq.(\ref{eq2x}) the terms in 
the first sum ($\sim {\bf p}^2_i$) represent the kinetic energies of electrons, while other 
terms correspond to the electron-nuclear Coulomb attraction and/or electron-electron Coulomb 
repulsion. From this equation one finds the following relation for the expectation values (in 
atomic units):
\begin{eqnarray}
 \Bigl[\frac{N}{2} \langle {\bf p}^2_e \rangle - Q N \langle \frac{1}{r_{en}} \rangle 
 + \frac{N (N - 1)}{2} \langle \frac{1}{r_{ee}} \rangle \Bigr] = E \label{eq3x}
\end{eqnarray}
where the subscript $e$ means the electron, while the subscript $n$ stands for the nucleus.

Let us assume that we know the highly accurate solution of the Schr\"{o}dinger equation, i.e. 
the wave function $\Psi$ in Eq.(\ref{eq2x}) is known. In this case the following condition must 
be obeyed for these three expectation values:
\begin{eqnarray}
 \langle {\bf p}^2_e \rangle = Q \langle \frac{1}{r_{en}} \rangle 
 - \frac{(N - 1)}{2} \langle \frac{1}{r_{ee}} \rangle \label{eq4x}
\end{eqnarray}
which is known as the virial theorem (see, e.g., \cite{Fock}). Now, after a few steps of 
simple transformations one finds the following formula for the total energy $E$ (in atomic 
units)
\begin{eqnarray}
 E &=& - \frac12 N \cdot Q \cdot \langle \frac{1}{r_{en}} \rangle + \frac14 N (N - 1) \cdot 
 \langle \frac{1}{r_{ee}} \rangle \nonumber \\ 
 &=& - \frac12 N \cdot Q \cdot f_1(N,Q) + \frac14 N (N - 1) \cdot f_2(N,Q) \label{eq5x}
\end{eqnarray}
where $f_1(N,Q)$ and $f_2(N,Q)$ are the functions of the number of electrons $N$ and 
electric charge of the nucleus $Q$. For neutral atoms we have $Q = N$, while for positively 
charged ions $Q > N$. For instance for the two-electron C$^{4+}$ ion (i.e. for $N$ = 2 and 
$Q$ = 6) we found that $\langle \frac{1}{r_{en}} \rangle \approx$ 5.687615325399107305988929 
$a.u.$ and $\langle \frac{1}{r_{ee}} \rangle \approx$ 3.438890700992227050791016158 $a.u.$ 
(results from calculations with 3700 basis functions). By using these expectation values one 
finds the total energy $E$ from Table IV. In the general case, $\langle \frac{1}{r_{ee}} 
\rangle = \lambda(N,Q)  \langle \frac{1}{r_{en}} \rangle$, where the paramter $\lambda < 1$.
The factor $\lambda(N,Q)$ can be approximated to very high accuracy with the use of highly 
accurate results obtained for different $N$ and $Q$. 

The formula, Eq.(\ref{eq5x}), is also applied in the case of $N = 1$ (hydrogen-like atoms and 
ions). In this case the electron-electron repulsion is not defined, but it is multiplied by a
factor which equals zero. As follows from Eq.(\ref{eq5x}), if we can determine the two unknown 
functions $f_1(N,Q)$ and $f_2(N,Q)$ in Eq.(\ref{eq5x}), then we can predict the total energy 
of an arbitrary $N-$electron atom/ion to very high accuracy. It can be done, e.g., by using 
the results of highly accurate computations of different few-electron atoms and ions (with 
different $Q$ and $N$). If we know the results of highly accurate computations for a large
number of light atoms/ions, then we can reconstruct the two functions $f_1(N,Q)$ and 
$f_2(N,Q)$ in Eq.(\ref{eq5x}) to very high accuracy. This procedure works well for mixtures 
of light elements at different temperatures and allows one to predict the properties of 
high-temperature plasmas which consists of such elements.

\newpage
\begin{table}[tbp]
   \caption{The total energies $E$ and expectation values of the electron-nuclear delta-function
            $\delta_{eN}$, electron-nuclear cusp $\nu_{eN}$ and some other operators for the 
            two-electron lithium ion Li$^{+}$ (in atomic units). $K$ is the total number of 
            basis functions used.}
     \begin{center}
     \begin{tabular}{cccccc}
      \hline\hline
 $K$ & $E$(Li$^{+}$) & & $\langle \delta_{eN} \rangle$ &  & $\nu_{eN}$ \\
     \hline
 3500 & -7.279913 412669 305964 91708 & & 6.8520 094343 431 & & -3.0000 00000 158 \\
 
 3700 & -7.279913 412669 305964 91743 & & 6.8520 094343 456 & & -3.0000 00000 125 \\

 3840 & -7.279913 412669 305964 91766 & & 6.8520 094343 460 & & -2.9999 99999 918 \\

 4000 & -7.279913 412669 305964 91785 & & 6.8520 094343 462 & & -2.9999 99999 901 \\
     \hline
 $K$ & $\frac12 \langle {\bf p}^2_1 \rangle$ & & $\langle {\bf p}_1 \cdot {\bf p}_2 \rangle$ & & $\frac12 \langle {\bf p}^2_N \rangle$ \\
      \hline\hline
 3500 & 3.63995 670633 465298 240 & & 0.288975 786393 989535 661 & & 7.56888 919906 329532 141 \\
 
 3700 & 3.63995 670633 465298 241 & & 0.288975 786393 989535 661 & & 7.56888 919906 329532 143 \\

 3840 & 3.63995 670633 465298 241 & & 0.288975 786393 989535 662 & & 7.56888 919906 329532 144 \\

 4000 & 3.63995 670633 465298 242 & & 0.288975 786393 989535 662 & & 7.56888 919906 329532 145 \\
    \hline\hline
  \end{tabular}
  \end{center}
  \end{table}
\begin{table}[tbp]
   \caption{The total energies $E$ and expectation values of the electron-nuclear delta-function
            $\delta_{eN}$, electron-nuclear cusp $\nu_{eN}$ and some other operators for the 
            two-electron berillium ion Be$^{2+}$ (in atomic units). $K$ is the total number of 
            basis functions used.}
     \begin{center}
     \begin{tabular}{cccccc}
      \hline\hline
 $K$ & $E$(Be$^{2+}$) & & $\langle \delta_{eN} \rangle$ & & $\nu_{eN}$ \\
     \hline
 3500 & -13.65556 623842 358670 20757 & & 17.1981 72544 645 & & -3.9999 99999 962 \\
 
 3700 & -13.65556 623842 358670 20767 & & 17.1981 72544 640 & & -3.9999 99999 921 \\

 3840 & -13.65556 623842 358670 20772 & & 17.1981 72544 638 & & -4.0000 00000 125 \\

 4000 & -13.65556 623842 358670 20777 & & 17.1981 72544 635 & & -4.0000 00000 148 \\
     \hline
 $K$ & $\frac12 \langle {\bf p}^2_1 \rangle$ & & $\langle {\bf p}_1 \cdot {\bf p}_2 \rangle$ & & $\frac12 \langle {\bf p}^2_N \rangle$ \\
      \hline\hline
 3500 & 6.82778 311921 179335 084 & & 0.420520 303439 441862 011 & & 14.07608 654186 302856 368 \\ 
 
 3700 & 6.82778 311921 179335 086 & & 0.420520 303439 441862 010 & & 14.07608 654186 302856 369 \\

 3840 & 6.82778 311921 179335 089 & & 0.420520 303439 441862 009 & & 14.07608 654186 302856 370 \\

 4000 & 6.82778 311921 179335 091 & & 0.420520 303439 441862 009 & & 14.07608 654186 302856 371 \\
    \hline\hline
  \end{tabular}
  \end{center}
  \end{table}
 \begin{table}[tbp]
   \caption{The total energies $E$ and expectation values of the electron-nuclear delta-function
            $\delta_{eN}$, electron-nuclear cusp $\nu_{eN}$ and some other operators for the 
            two-electron boron ion B$^{3+}$ (in atomic units). $K$ is the total number of basis 
            functions used.}
     \begin{center}
     \begin{tabular}{cccccc}
      \hline\hline
  $K$ & $E$(B$^{3+}$) & & $\langle \delta_{eN} \rangle$ & & $\nu_{eN}$ \\
     \hline\hline
 3500 & -22.03097 1580242 781541 65339 & & 34.758 743660 955 & & -5.0000 0000 319 \\
 
 3700 & -22.03097 1580242 781541 65376 & & 34.758 743660 965 & & -5.0000 0000 235 \\

 3840 & -22.03097 1580242 781541 65394 & & 34.758 743660 947 & & -5.0000 0000 107 \\

 4000 & -22.03097 1580242 781541 65418 & & 34.758 743660 935 & & -5.0000 0000 119 \\
      \hline
 $K$ & $\frac12 \langle {\bf p}^2_1 \rangle$ & & $\langle {\bf p}_1 \cdot {\bf p}_2 \rangle$ & & $\frac12 \langle {\bf p}^2_N \rangle$ \\
      \hline\hline
 3500 & 11.01548 579012 139077 100 & & 0.552752 631642 101467 789 & & 22.58372 421188 488300 979 \\ 
 
 3700 & 11.01548 579012 139077 089 & & 0.552752 631642 101467 734 & & 22.58372 421188 488300 952 \\

 3840 & 11.01548 579012 139077 086 & & 0.552752 631642 101467 715 & & 22.58372 421188 488300 942 \\

 4000 & 11.01548 579012 139077 083 & & 0.552752 631642 101467 701 & & 22.58372 421188 488300 938 \\
     \hline\hline
  \end{tabular}
  \end{center}
  \end{table}
  \begin{table}[tbp]
   \caption{The total energies $E$ and expectation values of the electron-nuclear delta-function
            $\delta_{eN}$, electron-nuclear cusp $\nu_{eN}$ and some other operators for the 
            two-electron carbon ion C$^{4+}$ (in atomic units). $K$ is the total number of basis 
            functions used.}
     \begin{center}
     \begin{tabular}{cccccc}
      \hline\hline
 $K$ & $E$(C$^{4+}$) & & $\langle \delta_{eN} \rangle$ & & $\nu_{eN}$ \\
     \hline\hline
 3500 & -32.40624 660189 853031 05527 & & 61.443 578056 445 & & -5.9999 99998 765 \\
 
 3700 & -32.40624 660189 853031 05535 & & 61.443 578056 514 & & -5.9999 99999 871 \\

 3840 & -32.40624 660189 853031 05539 & & 61.443 578056 537 & & -6.0000 00000 048 \\ 

 4000 & -32.40624 660189 853031 05542 & & 61.443 578056 543 & & -6.0000 00000 037 \\ 
      \hline
 $K$ & $\frac12 \langle {\bf p}^2_1 \rangle$ & & $\langle {\bf p}_1 \cdot {\bf p}_2 \rangle$ & & $\frac12 \langle {\bf p}^2_N \rangle$ \\
      \hline\hline
 3500 & 16.20312 330094 926515 523 & & 0.685334 822135 598924 527 & & 33.09158 142403 412923 500 \\ 
 
 3700 & 16.20312 330094 926515 524 & & 0.685334 822135 598924 535 & & 33.09158 142403 412923 502 \\

 3840 & 16.20312 330094 926515 525 & & 0.685334 822135 598924 536 & & 33.09158 142403 412923 502 \\

 4000 & 16.20312 330094 926515 525 & & 0.685334 822135 598924 537 & & 33.09158 142403 412923 503 \\
     \hline\hline
  \end{tabular}
  \end{center}
  \end{table}
  \begin{table}[tbp]
   \caption{The nuclear radius $R$ ($fm$), parameter $b$, factors $X$ and $Y$ (see the main 
            text) and field components of the total isotopic shift $\Delta E^{fs}$ (all values 
            are in atomic units) for each isotope.}
     \begin{center}
     \begin{tabular}{ccccccc}
      \hline\hline
 isotope & $Q$ & $R$ & $b$ & X & Y & $\Delta E^{fs}$ \\
     \hline\hline 
 ${}^{6}$Li  & 3 & 2.5385 & 0.99976034018621 & 4.297056149289$\cdot 10^{-8}$ & 0.81154478 & 2.389469390$\cdot 10^{-7}$ \\

 ${}^{7}$Li  & 3 & 2.4312 & 0.99976034018621 & 4.297056149289$\cdot 10^{-8}$ & 0.74440369 & 2.191782729$\cdot 10^{-7}$ \\
       \hline
 ${}^{9}$Be  & 4 & 2.5180 & 0.99957389838248 & 5.749782211793$\cdot 10^{-8}$ & 0.79852685 & 7.896292299$\cdot 10^{-7}$ \\
      \hline
 ${}^{10}$B  & 5 & 2.4278 & 0.99933413638122 & 7.219245621776$\cdot 10^{-8}$ & 0.74241787 & 1.862963331$\cdot 10^{-6}$ \\

 ${}^{11}$B  & 5 & 2.4059 & 0.99933413638122 & 7.219245621776$\cdot 10^{-8}$ & 0.72909310 & 1.829527229$\cdot 10^{-6}$ \\
       \hline
 ${}^{12}$C & 6 & 2.4073 & 0.99904101579314 & 8.7092766851788$\cdot 10^{-8}$ & 0.76867983 & 4.113429628$\cdot 10^{-6}$ \\

 ${}^{13}$C & 6 & 2.4614 & 0.99904101579314 & 8.7092766851788$\cdot 10^{-8}$ & 0.76315629 & 4.083871553$\cdot 10^{-6}$ \\

 ${}^{14}$C & 6 & 2.5037 & 0.99904101579314 & 8.7092766851788$\cdot 10^{-8}$ & 0.78958608 & 4.225305034$\cdot 10^{-6}$ \\
     \hline\hline
  \end{tabular}
  \end{center}
  \end{table}
\end{document}